\documentclass[draft]{agujournal2019}
\usepackage{natbib}
\usepackage{url} 
\usepackage{lineno}
\usepackage[inline]{trackchanges} 
\usepackage{soul}
\usepackage{amsmath} 
\draftfalse

\journalname{}

\begin{document}

\title{Do AI models predict storm impacts as accurately as physics-based models? A case study of the February 2020 storm series over the North Atlantic}


%
%


\authors{Hilla Afargan-Gerstman\affil{1}\thanks{}, Rachel W.-Y. Wu\affil{2},  Alice Ferrini\affil{2}, and Daniela I.V. Domeisen\affil{3,2} }
 
\affiliation{1}{Oeschger Centre for Climate Change Research, Institute of Geography, University of Bern, Switzerland}
\affiliation{2}{Institute for Atmospheric and Climate Science, ETH Zurich, Zurich, Switzerland}
\affiliation{3}{Faculty of Geosciences and Environment, University of Lausanne, Lausanne, Switzerland}


\correspondingauthor{Hilla Afargan-Gerstman}{hilla.gerstman@unibe.ch}

\begin{abstract}

The emergence of data-driven weather forecast models provides great promise for producing faster, computationally cheaper weather forecasts, compared to physics-based numerical models. However, while the performance of artificial intelligence (AI) models have been evaluated primarily for average conditions and single extreme weather events, less is known about their capability to capture sequences of extreme events, states that are usually accompanied by multiple hazards. The storm series in February 2020 provides a prime example to evaluate the performance of AI models for storm impacts. This event was associated with high surface impacts including intense surface wind speeds and heavy precipitation, amplified regionally due to the close succession of three extratropical storms.  In this study, we compare the performance of data-driven models to physics-based models in forecasting the February 2020 storm series over the United Kingdom. We show that on weekly timescales, AI models tend to outperform the numerical model in predicting mean sea level pressure (MSLP), and, to a lesser extent, surface winds. Nevertheless, certain ensemble members within the physics-based forecast system can perform as well as, or occasionally outperform, the AI models. Moreover, weaker error correlations between atmospheric variables suggest that AI models may overlook physical constraints.
This analysis helps to identify gaps and limitations in the ability of data-driven models to be used for impact warnings, and emphasizes the need to integrate such models with physics-based approaches for reliable impact forecasting.
\end{abstract}


 




\section{Introduction}

The recent emergence of artificial intelligence (AI) provides new pathways for producing weather forecasts in less time and at lower computational cost \citep{rasp2024weatherbench, molina2023review}.  However, there is a need to evaluate the capability of these models in reproducing extreme events and their associated surface impacts at the regional scale, especially in the context of weather extremes that occur in close succession, for instance recurrence of extratropical storms, which often leads to compounding effects of damaging winds and flooding.


Data-driven forecasts have been compared to physics-based forecasts for single storm events \citep{charlton2024ai,pasche2025validating}. \citet{charlton2024ai} analyzed this aspect by comparing forecasts based on data-driven versus physics-based NWP models for the cyclone Ciarán, which hit Europe in November 2023. They found that data-driven forecasts, despite accurately reproducing the synoptic-scale structure of the storm, failed to accurately estimate the wind speed.
For the 2021 North American winter storm, \citet{pasche2025validating} found that the data-driven models performed comparably or better than the physics-based forecast, particularly in forecasting compound winter storm conditions (including wind speed, temperature and wind chill). However, they emphasized that this result may be event-specific, and therefore a broader, systematic validation of data-driven forecast is needed across diverse types of extremes and impact metrics.

Other studies have evaluated the performance of data-driven weather prediction models in forecasting weather extremes 10 days in advance \citep{pasche2025validating,olivetti2024data}. Generally, data-driven models were found to perform as good as the deterministic forecast of the physics-based model of the European Centre for Medium-Range Weather Forecasts (ECMWF) for near-surface temperature and wind extremes, however their performance varies by region and forecast lead time. Furthermore, while data-driven models can reach similar accuracy to ECMWF's NWP model at the local scale, their performance is lower when variables are aggregated over space and time \citep{pasche2025validating}.


However, despite recent progress in forecasting of single extreme weather events, the occurrence of multiple weather events, in which an accurate prediction of the sequence of events is crucial for assessing their potentially devastating impacts, has received less attention in the literature. Storm clustering events provide an opportunity to evaluate the predictability of a sequence of weather systems, rather than focusing on a single event. Extratropical cyclone clustering in the North Atlantic is associated with strong winds and large amounts of precipitation affecting the same area within a short time span, posing an increased risk to physical infrastructure and human lives over Europe \citep{dacre2020serial, pinto2013serial, Priestley2020}. 


Such a close succession of storms occurred in February 2020, when the three cyclones Ciara, Dennis, and Jorge hit the United Kingdom within a short time period. 
Such conditions tend to result in a shorter recovery time between the events, often leading to serious socioeconomic consequences such as flooding of rivers, disruption of transportation and damage to infrastructure.

On February 8, Storm Ciara (also known as Sabine, Elsa) hit the United Kingdom (UK), bringing windy weather with persistent heavy rain, especially over northwestern England. A week later, on February 15, Storm Dennis (also known as Victoria), one of the deepest Atlantic depressions on record \citep{davies2021wet}, impacted the British Isles and north-western Europe and brought even wetter conditions, prompting the UK Met Office to issue red weather warnings for part of South Wales. 
While wind speed and tidal surges during storms Ciara and Dennis were substantially higher than the February mean, precipitation exhibited the largest anomaly, leading to extensive impacts over Western Britain \citep{jardine2023multidisciplinary}.
Finally, on February 28, Storm Jorge hit the UK, which, although the least intense of the three, 
Storm Jorge added to the extremely prolonged period of rainfall \citep{griffin2025putting,sefton20212019}) and contributed to worsening the overall damage already caused by Ciara and Dennis. 

Overall, the meteorological conditions in February 2020 led to exceptionally heavy rainfall across the United Kingdom, making it the wettest February on record for many regions \citep{davies2021wet,griffin2025putting,sefton20212019}. River flows responded rapidly to this rainfall, as soils were already near saturation from preceding precipitation events, resulting in record river discharges and extended flooding. The following river floods caused severe damage across Wales, northern England, and the Midlands. The total insured losses from the February 2020 UK floods were evaluated at approximately GBP 368 million \citep{PERILS2021UKFloods}. 
Specifically, the total industry loss for storms Ciara and  Dennis (including damage over the British Isles and Continental Europe) were estimated at EUR 1,571 million and EUR 350 million, respectively \citep{PERILS2021Sabine,PERILS2021Dennis}.  

Data-driven forecasts have demonstrated improved performance on short-term predictions \citep{leinonen2023thunderstorm,andrychowicz2023deep} as well as medium range forecasts (up to 2 weeks in advance) for various atmospheric variables (including temperature and wind) and their extremes \citep{lam2023learning,price2023gencast,rasp2024weatherbench,nguyen2023climax,pasche2025validating,olivetti2024data, zhang2025numerical}.
However, forecasting a series of storms, rather than a single event, on weekly timescales can be a challenging task \citep{dacre2020serial}. 
On these timescales, extratropical cyclone clustering may depend on the properties of the primary cyclone and the conditions in which it develops, while remote drivers, such as sea surface temperature anomalies and stratospheric variability, can modulate the development of storms and their propagation. Specifically,  unusually strong stratospheric polar vortex conditions may increase the likelihood of intense extratropical cyclones impacting the UK \citep{afargan2025winter}, which was also shown for the storm clustering over the UK in February 2022 \citep{williams2025strong}. 
Furthermore, the compounding effect of multiple, consecutive extreme events is important for accurate prediction of their local impacts. Assessing model performance for case-studies that can lead to substantial surface impact when temporally and spatially aggregated is a critical step towards increased reliability of impact predictions by data-driven models and advancing their potential use for socioeconomic preparedness and early warning. 

This study investigates the performance of data-driven weather prediction models in capturing the dynamics of extratropical storm activity over the North Atlantic and Europe through three representative case studies of a rapid succession of extratropical storms. Specifically, we compare the ability of such models to represent storm clustering and forecast error correlations between physically-linked variables against a state-of-the-art dynamical weather prediction models.

\section{Methods} 

In this study, we evaluate the performance of medium-range forecasts by a physics-based model and two data-driven model against reanalysis data. 
All datasets are obtained from WeatherBench 2, an open source evaluation framework for medium-range global weather forecasting that provides 
datasets on Google Cloud Storage with a time step of 6 hours at a resolution of 0.25° or higher \citep{rasp2024weatherbench}. In this study, we use 6-hourly data at 1.5$^\circ$ spatial resolution, averaged to a daily mean for specific analyses. We focus the analysis on the forecasts initialized at 00 UTC.

We focus on three different initialization dates: February 1, 8, and 21 (referred to as the ''forecast initialization date''). For each initialization date, the forecast is validated for a window up to a lead time of 10 days. Lead time is defined as the time interval between the initialization day and the day for which the forecast is validated (''valid time''). The initialization dates are selected such that the  peak of storm intensity of the storms of interest (Ciara, Dennis and Jorge) occurs on the 8th lead day of each forecast. All forecasts are validated against ERA5 reanalysis \citep{Hersbach2020}.





\subsection{Numerical Weather Prediction Models}

Numerical weather prediction (NWP) models solve sets of mathematical equations for the atmosphere and oceans to create a prediction of the weather based on current weather conditions. Here we use 50-member operational ECMWF IFS ensemble forecasts generated by the Integrated Forecasting System (IFS) produced by the European Centre for Medium-Range Weather Forecasts (ECMWF). These  ensemble forecasts, were retrieved from the THORPEX Interactive Grand Global Ensemble (TIGGE) archive \citep{bougeault2010thorpex} 
through WeatherBench 2. 
IFS ensemble mean (IFS ENS mean) is computed by taking an average over the 50 members and is used as a baseline in \citet{rasp2020weatherbench} as it performs well on deterministic error metrics. 


\subsection{Data-driven Models}

With the recent advancements in artificial intelligence, it has been possible to expand the number of forecast datasets by including data-driven weather forecasting models. Here we evaluate two data-driven models: 
Pangu-Weather \citep{Bi2023pangu}, developed by Huawei and based on a three-dimensional Earth-specific transformer and hierarchical temporal aggregation; and
GraphCast \citep{lam2023learning}, developed by Google DeepMind and based on graph neural networks.

Data-driven models typically follow a three-phase development process: training, validation, and prediction (testing). Specifically, Pangu-Weather is trained in a first phase on past data, in this case, the ERA5 data from January 1979 to December 2017, validated for 2019 and  tested for the years 2018, 2020 and 2021. A similar process is used for Graphcast which, however, involves training on ERA5 data from January 1979 to December 2019.

\subsection{Model verification}




We evaluate the forecasts for two main variables that are directly relevant to storm impacts, namely mean sea level pressure (MSLP) and 10-m wind speed, which measure  the storm intensity and the associated impacts, respectively. These variables are averaged and evaluated over a fixed location over the UK (48 - 60$^\circ$N, 12$^\circ$W - 5$^\circ$E; Fig. \ref{fig:timeseries_fig1}) as the UK was heavily impacted by the successive passage of the storms in February 2020. Anomalies of MSLP and wind speed are computed as deviations from their daily climatological mean from 1990 to 2019, obtained from ERA5 reanalysis.

To quantify the error in forecasting cyclone intensity in terms of MSLP and wind speed, the Mean Error (ME) is calculated as:

\begin{linenomath*}
\begin{equation}
\text{ME} = \frac{1}{N} \sum_{i=1}^{N} (F_i - O_i)
\end{equation}
\end{linenomath*}

where $F_i$ denotes the forecasted value and, $O_i$ the observed value, both at  time i. The variable N represents the number of ensemble members. 
A perfect forecast would result in an ME of 0.

We also consider the Mean Absolute Error (MAE) for comparing the average magnitude of the forecast errors, regardless of their sign (i.e. overestimation or underestimation).



\section{Results}

In this section, we present a comparison between the physics-based  weather prediction model and the data-driven models, with aim of quantifying the models' ability to reproduce the characteristics of the observed storm clustering event, as well as their ability to maintain the correlations between physically-linked variables. We focus on the forecast verification of a series of extratropical storms over the UK in February 2020. 

\subsection{The storm series in February 2020} 

We analyze storms Ciara, Dennis and Jorge, which hit the UK in rapid succession in February 2020. Although this event is not defined as a storm clustering events in the literature \citep{davies2021wet}, it nevertheless featured a close temporal succession of the three storms over the northern part of the North Atlantic and the UK, with intensity peaks on 8-9 February for storm Ciara, 15-16 February for storm Dennis and 28-29 February for storm Jorge. Similar close succussions of cyclones over Western Europe were recorded in several past winters, including 1990, 1993, 1999, 2007 and 2014 \citep[e.g.,][]{klawa2003model,fink2009european,dacre2020serial}. 

\begin{figure}[h]
    \centering
    \includegraphics[width=0.8\textwidth]{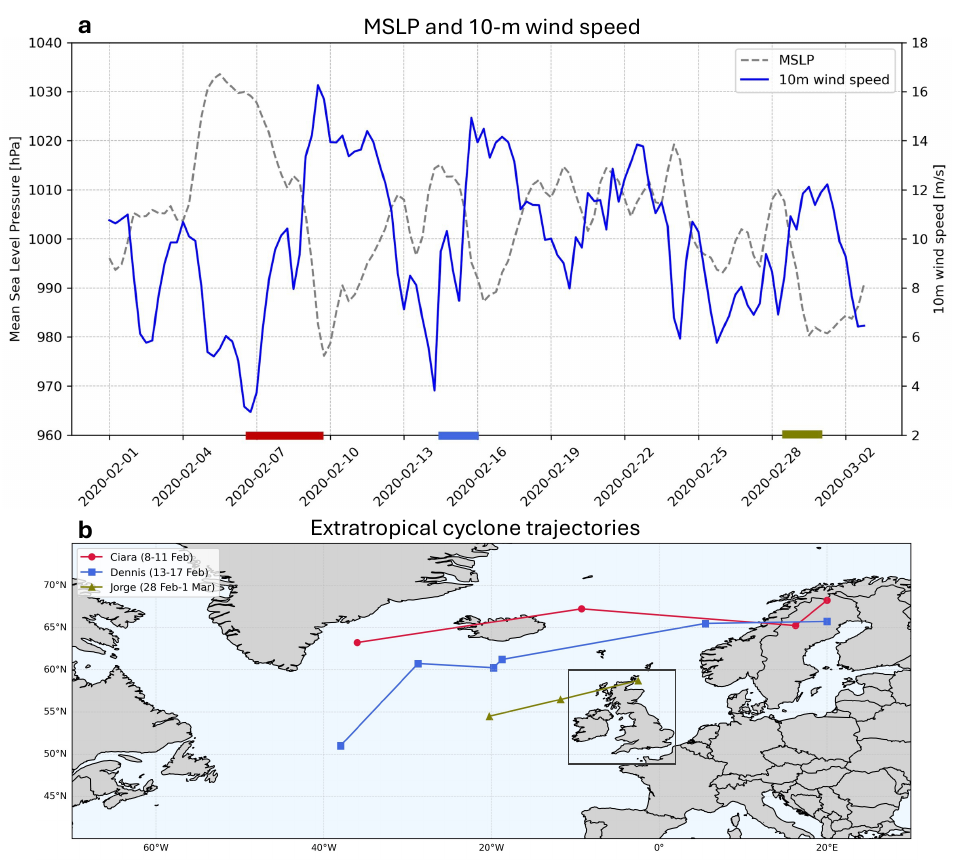} 
    \caption{(a) Time series of MSLP (dashed grey line) and 10-meter wind speed (solid blue line) over the UK (12$^\circ$W-5$^\circ$E, 48$^\circ$N-60$^\circ$N) during the storm series in February 2022. (b) Trajectories of the three storms (Ciara, Dennis and Jorge) over the North Atlantic and Western Europe, based on daily minimum MSLP anomalies computed relative to daily 30-year climatology (see the Methods section for details). 
    }
    \label{fig:timeseries_fig1}
\end{figure}

As a first qualitative assessment, we analyze the timeseries of MSLP (dashed) and 10m wind speed (in blue) observed over the UK region during February 2020 (Fig. \ref{fig:timeseries_fig1}a), and their corresponding cyclone trajectories in the North Atlantic basin: Ciara (in red), Dennis (in blue), and Jorge (in green) (Fig. \ref{fig:timeseries_fig1}b). 
Storm Ciara developed rapidly on February 7 (but was
particularly noteworthy already on February 4) reaching its maximum intensity one day later between Iceland and Greenland, with a strongly negative minimum MSLP anomaly of -60 hPa, associated with a wind speed peak of +15 m/s (Fig. \ref{fig:timeseries_fig1}a). The cyclonic conditions brought intense northwesterly winds towards the UK, recording one of the highest wind anomalies of the month in the region (+10 m/s), accompanied by a drop in MSLP (-45 hPa). The cyclone then gradually weakened over the following days as it moved towards Scandinavia. Storm Ciara was characterized by an exceptionally wide area that was affected by damaging wind gusts across the British Isles and Continental Europe \citep{PERILS2021Sabine}.

A second cyclone, Storm Dennis, forms on 15 February in the middle North Atlantic and rapidly intensifies as it moves northeastward. It reaches its peak intensity on February 16, recording the most intense pressure drop of the month, with a MSLP anomaly of -65 hPa and a maximum wind speed anomaly of +18 m/s (Fig. \ref{fig:timeseries_fig1}a). The winds associated with Dennis extend across a large area of the North Atlantic and Western Europe, significantly affecting the UK, where maximum wind speed anomalies (+11 m/s) and the strongest negative minimum MSLP anomaly (-55 hPa) of the month in this region are recorded. In the following days, Dennis gradually loses intensity as it approaches Norway, dissipating completely by February 20. Storm Jorge developed on 28 February to the west of the UK (Fig. \ref{fig:timeseries_fig1}a) and within a few hours reaches a MSLP anomaly of -50 hPa and a wind speed anomaly of +13 m/s northwest of Ireland. Despite being the least intense of the three cyclones over the North Atlantic, Jorge had comparably strong UK wind anomalies. Finally, the storm weakens and dissipates in the first days of March.

\begin{figure}[h]
    \centering    \includegraphics[width=\textwidth]{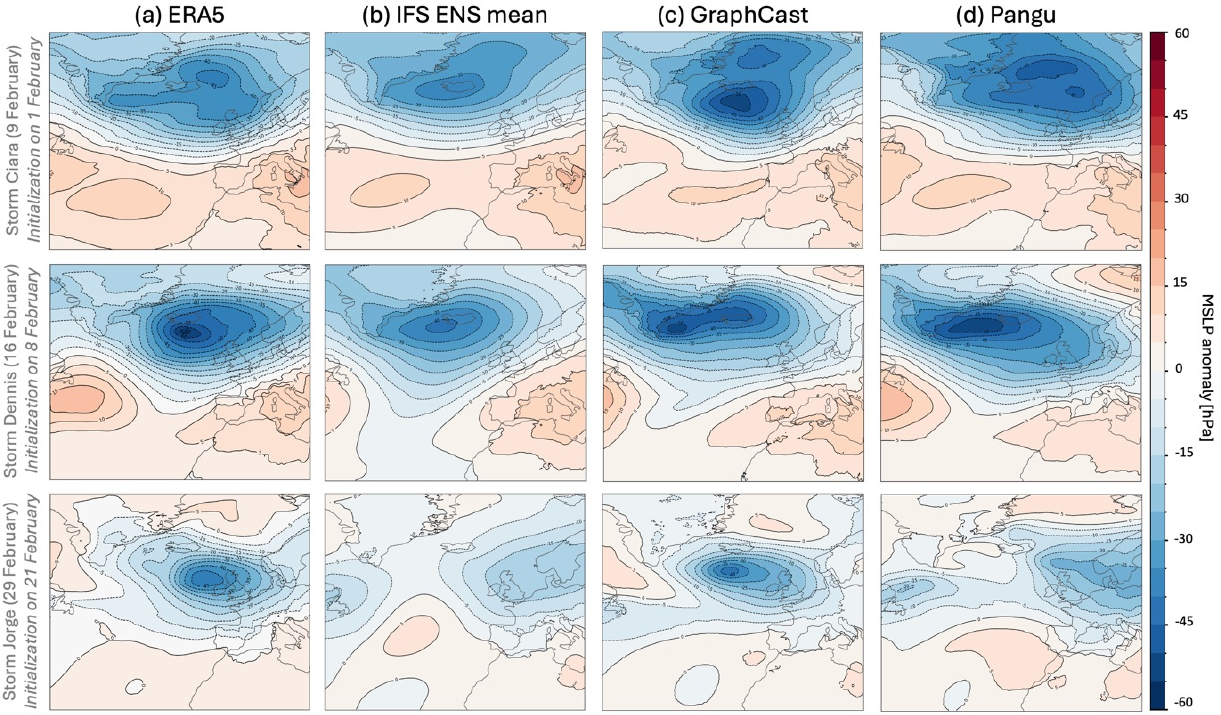} 
    \caption{MSLP anomalies (shading) over the North Atlantic and Western Europe (20 - 80$^\circ$N, 60$^\circ$W - 20$^\circ$E) for the days of peak intensity on 9 February (upper row), 16 February (middle row) and 29 February (bottom row). Data are derived from (a) ERA5 reanalysis, (b) IFS ENS mean, (c) GraphCast, and (d) Pangu-Weather weather. Anomalies are computed relative to the 1990–2019 daily climatology in ERA5. }
    \label{fig:mslp_anom_all}
\end{figure}

\begin{figure}[h]
    \centering    \includegraphics[width=\textwidth]{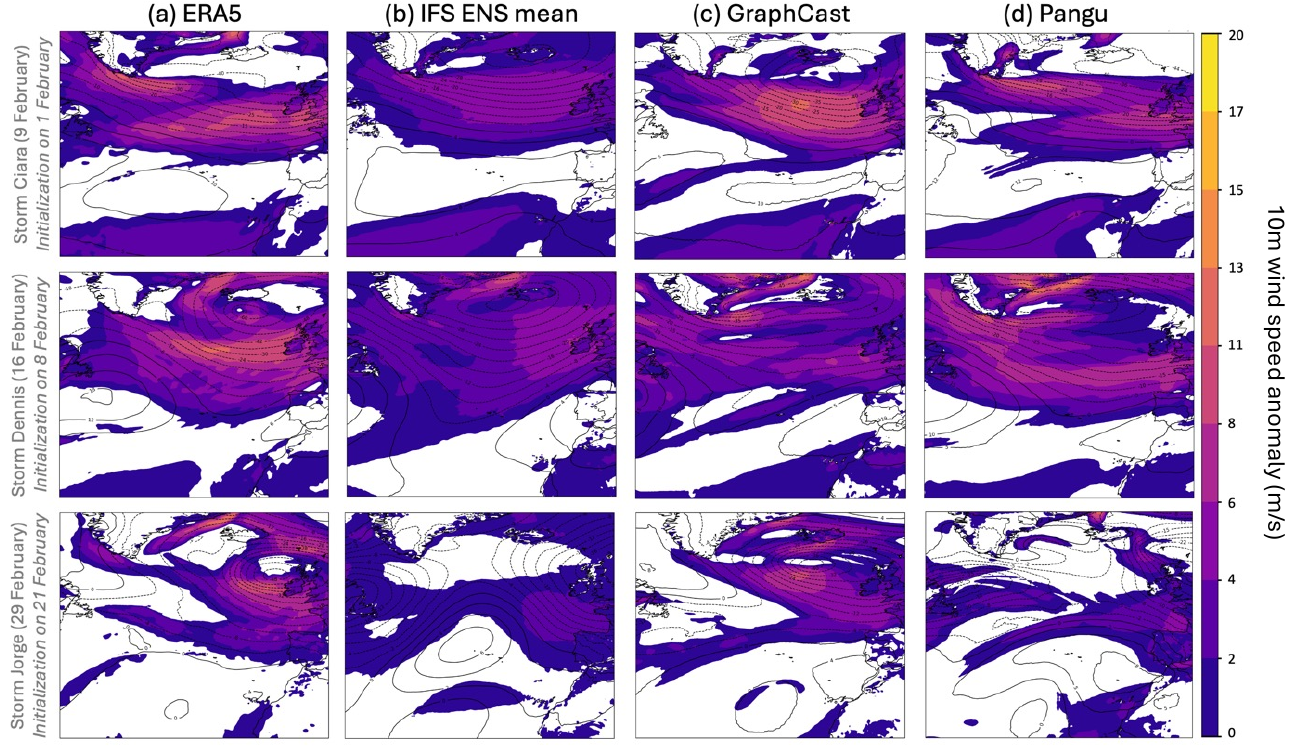} 
    \caption{Same as Fig. \ref{fig:wind_anom_all}, but for 10m wind speed (shading).}
    \label{fig:wind_anom_all}
\end{figure}

\subsection{Forecast verification} 

Figure \ref{fig:mslp_anom_all} and Figure \ref{fig:wind_anom_all} show MSLP anomalies (shading) and 10-m wind anomalies (shading), respectively, for the three storms (Ciara, Dennis, Jorge) on their day of maximum development, comparing reanalysis (ERA5, column a) with NWP model forecasts (IFS ENS mean, column b) and the two data-driven models (GraphCast and Pangu-Weather, columns c and d), at lead times of 8 days with respect to the forecast initialization date (see Figure labels for the dates). 

Overall, all models are able to capture realistic storm structures at their peak magnitudes (Fig. \ref{fig:mslp_anom_all} and \ref{fig:wind_anom_all}). Data-driven models (columns c,d of Fig. \ref{fig:mslp_anom_all} and \ref{fig:wind_anom_all}) predict magnitudes of MSLP and wind speeds that are more comparable to ERA5 than IFS ENS mean. The intensities of both MSLP and 10-m wind speed are clearly underestimated by IFS ENS mean (column b of Fig. \ref{fig:mslp_anom_all} and \ref{fig:wind_anom_all}).
However, these similarities in MSLP do not necessarily translate into an equally accurate forecast of the surface wind speed  (see also Table \ref{tab:bias_summary}). 

\begin{figure}
    \centering
    \includegraphics[width=\linewidth]{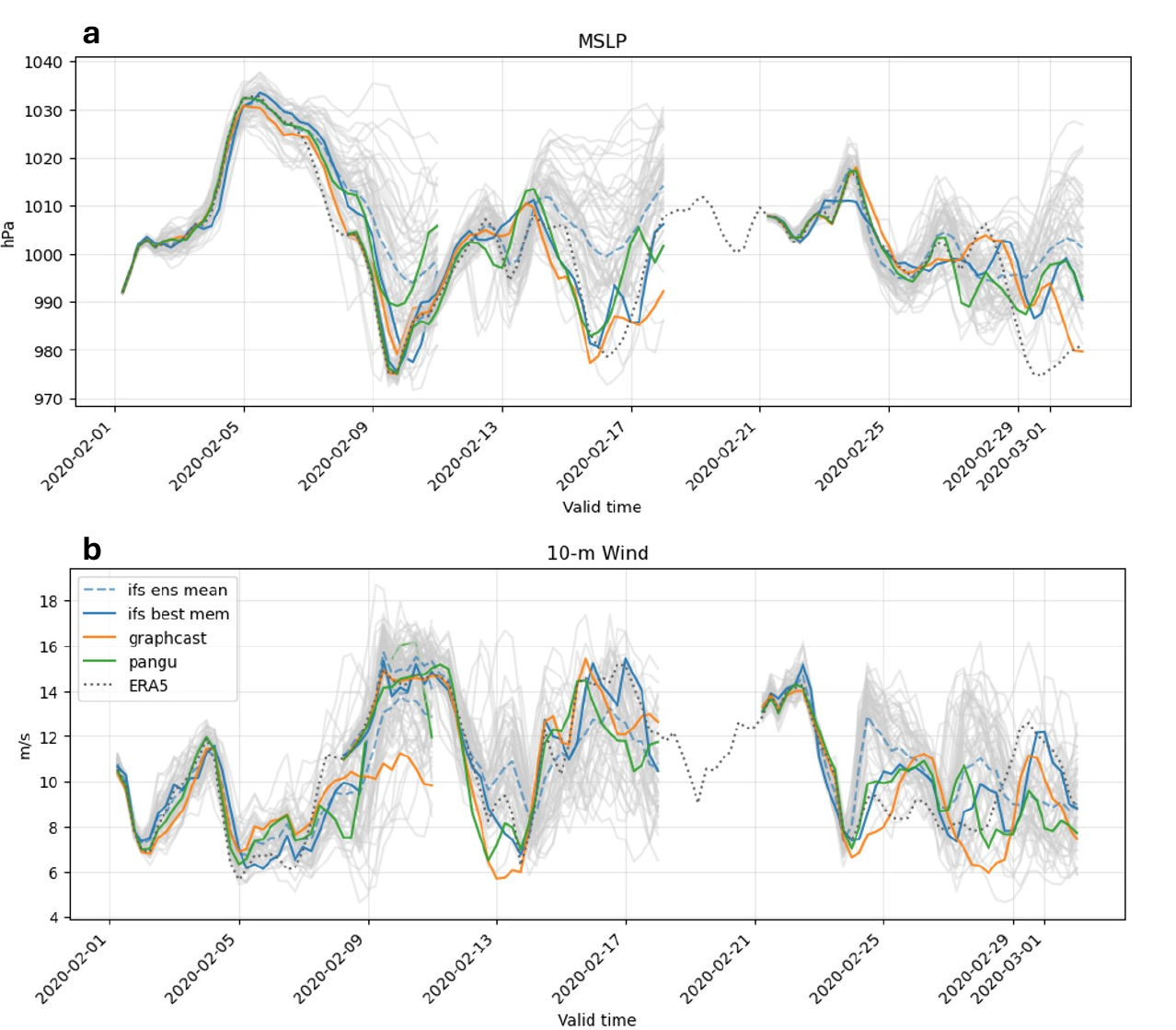}
    \caption{Predicted time series of (a) MSLP and (b) 10-m wind averaged over the UK for the forecast models for the three initialization dates: 1 February (for Storm Ciara), 8 February (Storm Dennis), and 21 February (Storm Jorge). Forecasts are plotted for three models: the physics-based model IFS (blue solid line) and two AI models: Graphcast (orange) and Pangu (green). IFS ensemble members are plotted in solid grey lines and the ensemble mean is plotted in dashed blue.  The best member of IFS in each initialization is highlighted in solid blue line. The ERA5 is plotted in dotted grey line. }
    \label{fig:wind_mslp_ifs_ens}
\end{figure}

Figure \ref{fig:wind_mslp_ifs_ens} further compares the evolution of MSLP (in hPa) and 10-m wind speed (in $m s^{-1}$) between the forecasts (colored lines) and against ERA5 reanalysis (dotted grey line) up to 10 days after initialization. 
At short lead times, all models produce MSLP and wind anomalies that closely match the observed values. 
However, as lead time increases, distinct differences emerge between the data-driven forecasts and the deterministic baseline of physics-based model. The IFS ENS mean (dashed blue line) captures the overall timing of each cyclone's deepening over the UK but simulate weaker cyclones, underestimating the MSLP minima of Ciara and Dennis by approximately 20 hPa and Jorge by about 25 hPa. As the cyclones decay, the MSLP in IFS ENS mean forecasts gradually weakens, following the observed dissipation rate. 
In contrast, the data-driven models (orange and green lines for GraphCast and Pangu-Weather, respectively) exhibit smaller amplitude variations in MSLP and a smoother temporal evolution of cyclone intensity and wind speed (Fig.~\ref{fig:wind_mslp_ifs_ens}). This leads to a more consistent representation across lead times and overall magnitudes closer to observations.

Yet, it is important to notice that while IFS ENS mean may underestimate the intensities of the cyclones and the associated surface wind, some IFS ENS ensemble members accurately predict the observed deepening and decay rates of the cyclone series (solid grey line). The best-performing member (solid blue line) is indicated in Fig.~\ref{fig:wind_mslp_ifs_ens} for each of the cyclone events. The best-performing member is the one with the smallest MAE in 10-m wind speed, averaged over all lead times. The evolution of the best-performing member is found to be similar to that of the data-driven models.  

\begin{table}[ht]
\centering
\caption{Mean absolute error (MAE) for MSLP and 10-m surface wind, averaged over the UK, for all storm events and all models. MAE is computed using 6-hourly forecast data initialized at 00:00~UTC. 'IFS EM' denotes the ensemble mean, and 'IFS best member' refers to the best-performing ensemble member for a 10-day forecast. The rightmost column represents the average of IFS EM, GraphCast, and Pangu-Weather.}
\begin{tabular}{lccccc}
\hline
\textbf{Storm} & \textbf{IFS EM} & \textbf{IFS best member} & \textbf{GraphCast} & \textbf{Pangu-Weather} & \textbf{Mean} \\
\hline
\multicolumn{6}{c}{\textbf{MSLP MAE [hPa]}} \\
Storm Ciara  & 5.46 & 4.24 & 2.07 & 4.40 & 3.98 \\
Storm Dennis & 5.52 & 3.59 & 4.15 & 4.40 & 4.69 \\
Storm Jorge  & 6.83 & 5.55 & 3.45 & 5.20 & 5.16 \\
\textbf{Mean (all storms)} & \textbf{5.94} & \textbf{4.46} & \textbf{3.23} & \textbf{4.66} & \textbf{4.61} \\
\hline
\multicolumn{6}{c}{\textbf{Wind MAE [m/s]}} \\
Storm Ciara  & 0.99 & 0.77 & 1.57 & 1.00 & 1.19 \\
Storm Dennis & 1.09 & 0.64 & 0.87 & 1.01 & 0.99 \\
Storm Jorge  & 1.61 & 1.02 & 1.26 & 1.36 & 1.41 \\
\textbf{Mean (all storms)} & \textbf{1.23} & \textbf{0.81} & \textbf{1.24} & \textbf{1.12} & \textbf{1.20} \\
\hline
\end{tabular}
\label{tab:bias_summary}
\end{table}

Table ~\ref{tab:bias_summary} summarizes the storm forecast verification for MSLP and surface wind over the UK, by computing the Mean Absolute Error for each storm event separately. The best-performing member of IFS has lower MAE than IFS ENS mean in both MSLP and wind. The best member of IFS is also most skillful in predicting the 10-m wind among all models with the lowest average MAE of 0.81 $m s^{-1}$, followed by Pangu-Weather, GraphCast and IFS ENS mean. In terms of MSLP, GraphCast is the most skillful model with averaged MAE of 3.23 hPa, followed by IFS best member, Pangu-Weather and IFS ENS mean. 

Overall, out of the three storms of the February 2020 cluster, storm Ciara is the best-predicted storm by the models in MSLP with averaged MAE of 3.98 hPa. In terms of surface wind speed, storm Dennis is best predicted with MAE of 0.99 $m s^{-1}$ averaged over the models.

\subsection{Sources of forecast bias} 

\begin{figure}[h]
    \centering
    \includegraphics[width=\textwidth]{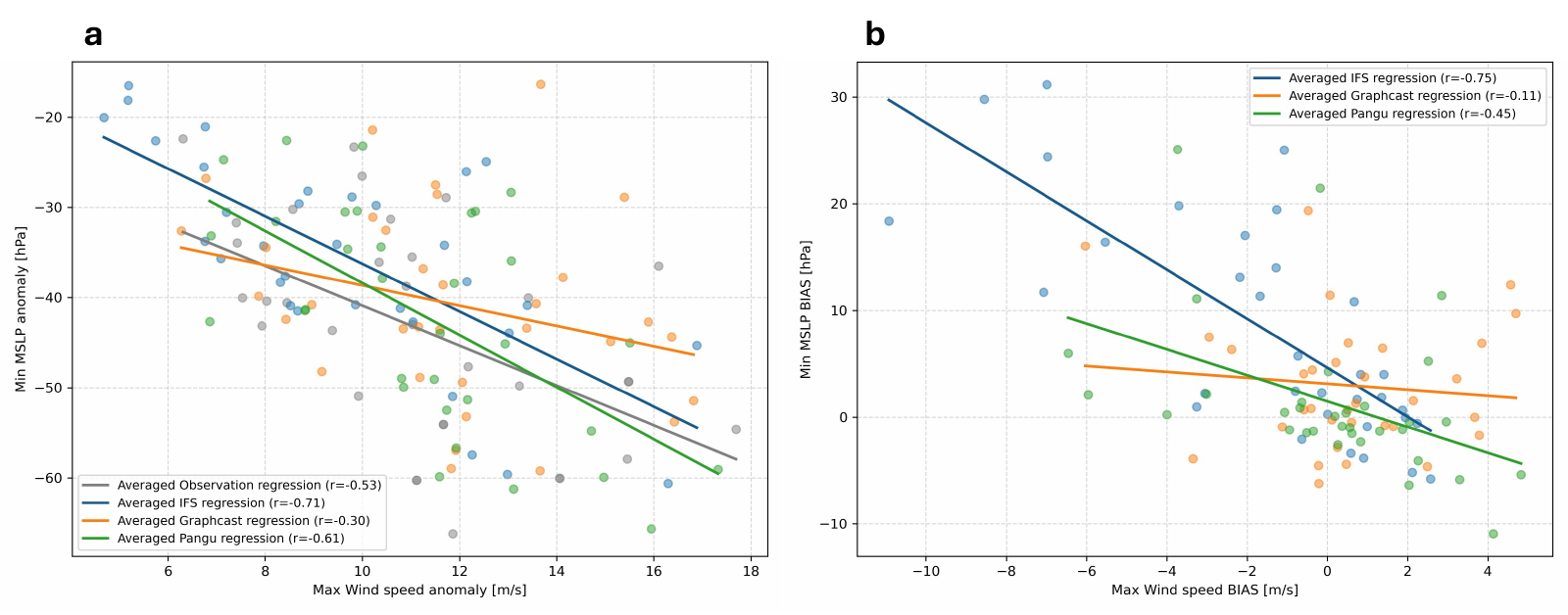} 
    \caption{(a) Scatter plot of daily values of maximum 10m wind speed anomaly (averaged over the UK; Fig \ref{fig:timeseries_fig1}) and minimum MSLP anomaly (averaged over the Euro-Atlantic region), averaged across all initializations. Each dataset is represented by a different color: grey for ERA5 observations, blue for IFS ENS mean, orange for GraphCast, and green for Pangu-Weather. A linear regression line is fitted to each dataset, and the corresponding regression coefficient (r). (b) Same as panel (a), but for the relationship between maximum 10m wind speed bias  vs. minimum MSLP bias.}
    \label{fig:corr_mslp}
\end{figure}

Understanding the causes and sources of forecast bias in data-driven and dynamical models requires assessing the way these models represent physical constrains.
For this purpose, we visualize the relationship between  storm intensity, measured as the minimum MSLP of each storm at each lead time of the forecast, and the associated maximum surface wind speed (Fig. \ref{fig:corr_mslp}a), as well as their respective errors (Fig. \ref{fig:corr_mslp}b). 

Fig.~\ref{fig:corr_mslp}a shows the relationship between MSLP over the UK and the surface wind in the Euro-Atlantic region. 
For each dataset, a regression line is then fitted  to quantify the strength and sign (positive or negative) of the relationship between these two variables, and the corresponding correlation coefficient (r) is calculated.  

Each point in Fig.~\ref{fig:corr_mslp}a indicates the individual days from all initializations, and the colors are used to distinguish the different forecast models and observations. Each dataset is associated with its respective regression line and regression coefficient (r). Overall, a negative correlation is observed, meaning that low MSLP values are associated with high wind speed values, and the more negative this value is, the stronger the negative correlation. IFS ENS mean exhibits a strong regression coefficient (r=-0.71) between the maximum wind speed anomalies and the minimum MSLP anomalies, while the data-driven models show a weaker relationship, especially for GraphCast (r=-0.30) (Fig.~\ref{fig:corr_mslp}a). Compared to the observations, which exhibit a regression coefficient of r=-0.53, IFS shows a stronger inverse relationship between MSLP and wind speed, potentially overestimating this correlation.

Next, we examine the error correlation for minimum MSLP and maximum surface wind in the forecast models for the three initializations (Fig.~\ref{fig:corr_mslp}b). Unlike physics-based models, AI models show weaker error correlations between physically linked variables, such as storm intensity and surface wind. Both GraphCast and Pangu-Weather show a weaker correlation compared to IFS, with r=0.11 and r=0.45, respectively. In comparison, IFS shows a relatively strong negative correlation, with r=0.71. 
These results indicate that for the AI models, storm intensity forecasting errors (as measured e.g., by maximum MSLP over the UK) do not necessarily lead to errors in surface wind prediction.

\section{Discussion and conclusions}

This study performs a comparison between physics-based and data-driven weather prediction models: ECMWF's IFS ENS mean, GraphCast, and Pangu-Weather, in forecasting a series of extratropical cyclones that hit the UK in February 2020.
Storm clustering is an extreme and compounding event often associated with substantial impacts due to the strong winds and increased risk of flooding. The ability of a weather model to predict the intensity of each storm in the cluster in terms of MSLP, wind speed, and their location with a high degree of accuracy is critical to support effective disaster preparedness, addressing an increasingly complex challenge due to their aggregated impacts \citep{afargan2025winter,williams2025strong}. In this context, the rapid evolution of the forecasting capabilities of AI models represents, on the one hand, a great opportunity for atmospheric sciences and, on the other, a significant challenge, as it is still unclear how well these models can reproduce extreme events. 

Our results show that both data-driven and physics-based models tend to underestimate storm intensity (MSLP) and surface wind intensity, particularly at lead times beyond 5-7 days. However, data-driven models (in this study, GraphCast and Pangu-Weather) show comparable or better skill as compared to the ensemble mean of the physics-based model in reproducing wind anomalies associated with extratropical cyclones (Table \ref{tab:bias_summary}). 
%

In addition, unlike the physics-based model, the data-driven models show weaker error correlations between physically linked variables, such as storm intensity (measured by the minimum MSLP of the storm) and surface wind, indicating an improved ability of data-driven models in predicting the surface wind field (and consequently, windstorm-related impacts), yet while potentially misrepresenting physical constraints. 
Hence, at current, predicting impacts from data-driven models remains challenging given the missing relationship between variables and biases. Specifically, the question arises if  an AI-based forecast of high surface wind speeds can be trusted for an impact-based warning if the associated storm is poorly resolved, misplaced, or incorrectly predicted in the data-driven model. 

At the same time our study demonstrates that while the ensemble mean of IFS overall has weaker prediction skill as compared to the data-driven models, there are single ensemble members within IFS that outperform the data-driven models. Therefore, given the deficient physical consistency between variables and biases in the data-driven models,  AI models (in their current state) can be recommended for use for impact warnings only with  access to a prediction from a physical model at the same time. In fact, given the high skill of single ensemble members, it might at current still be more advisable and lead to more trustworthiness to use data-driven methods to help select the best performing ensemble member from the physical model.  Ensemble sub-selection methods such as those outlined in \citep{dobrynin2018improved} and applied in \citep{famooss2025hybrid} for seasonal prediction based on simple statistical methods may potentially be applied to short- and extended-range forecasts using data-driven approaches for performing the sub-selection. Similar methods of combining physical and data-driven models exist within so-called hybrid forecasting methods for predicting impacts \citep[e.g.][]{materia2024artificial,slater2022hybrid}.

%

Overall, the results of this research highlight the significant potential and progress of data-driven models in improving the accuracy of predicting extreme events such as storm clustering. These findings are based on three representative case studies; thus, systematic evaluation is required to draw more general conclusions regarding the capability of data-driven models in forecasting extreme weather events, especially in terms of the physical consistency of extreme events and the associated impact predictions. In another study analyzing Storm Ciarán in November 2023 \citep{charlton2024ai}, in agreement the findings in our study, AI-driven models were found to capture the storm intensity evolution as successfully as the physics-based  models, although they underestimated the peak surface winds associated with the storm. 

%

In summary, combining  physics-based and data-driven models within hybrid modeling frameworks offers the potential to improve forecasts of extreme weather and climate impacts, including storm clustering events. Such advancements can support the development of more effective tools for local preparedness and early warning systems, thereby mitigating potentially devastating storm impacts.

\section*{Open Research Section}
The forecasts for all models are available through the WeatherBench 2 platform \citep{WB2_data_2024}. ERA5 reanalysis dataset \citep{Hersbach2020} is freely available through the Copernicus Climate Change Service \citep{CDS_ERA5_2024}, as well as through WeatherBench 2. 



\acknowledgments
This project has received funding from the Swiss National Science Foundation through project PZ00P2\_223676.  

\bibliography{bib_s2s_hila} 
%

\end{document}